# Combiner suivi de l'activité et partage d'expériences en apprentissage par projet pour les acteurs tuteurs et apprenants

Michel Christine [1], Lavoué Élise [2]

Christine.Michel@insa-lyon.fr, Elise.Lavoue@univ-lyon3.fr

[1] Université de Lyon, INSA Lyon, Laboratoire LIESP, F-69621 Villeurbanne, France
[2] Université Jean Moulin Lyon 3, IAE Lyon, Équipe MAGELLAN, Groupe SICOMOR, Lyon, France

**RÉSUMÉ :** Nos recherches ont pour objectif d'offrir des outils d'aide aux apprenants et tuteurs impliqués dans des activités d'apprentissage en projet en présence ou à distance. L'apprentissage par projet est souvent utilisé pour un apprentissage complexe (i.e. qui a pour but d'amener les étudiants à acquérir et développer diverses compétences et comportements). Ce type d'apprentissage repose sur le co-développement, la responsabilité collective et la coopération, les apprenants étant les principaux acteurs de leur apprentissage. Ces formations reposent sur des organisations riches et complexes, particulièrement pour les tuteurs et il est difficile de mettre en œuvre des stratégies pédagogiques innovantes. Notre propos, dans une démarche bottom-up, est (1) d'observer, selon des méthodes de Knowledge Management, une formation caractérisée par ces trois critères. La formation observée concerne l'apprentissage à la gestion de projet. Son observation nous permet de (2) mettre en évidence et analyser les problèmes vécus par les acteurs (étudiants, tuteurs, concepteurs) et de (3) proposer des outils capables de les résoudre ou les améliorer. Nous étudions en particulier la pertinence et les limites des outils existants concernant le suivi et le partage d'expérience. Nous proposons un résultat sous la forme de l'outil MEShaT (« *Monitoring and Experience Sharing Tool* ») et concluons sur les perspectives offertes par ces recherches.

**Mots clés** : Outils de suivi ; Pédagogie par projet ; Métacognition ; Web 2.0 ; Capitalisation des connaissances et savoir-faire ; Partage d'expérience.

**ABSTRACT:** Our work aims to study tools offered to students and tutors involved in face-to-face or blended project-based learning activities. Project-based learning is often applied in the case of complex learning (i.e. which aims at making learners acquire various linked skills or develop their behaviours). In comparison to traditional learning, this type of learning relies on co-development, collective responsibility and co-operation. Learners are the principal actors of their learning. These trainings rest on rich and complex organizations, particularly for tutors, and it is difficult to apply innovative educational strategies. Our aim, in a bottom-up approach, is (1) to observe, according to Knowledge Management methods, a course characterized by these three criteria. The observed course concerns project management learning. Its observation allows us (2) to highlight and to analyze the problems encountered by the actors (students, tutors, designers) and (3) to propose tools to solve or improve them. We particularly study the relevance and the limits of the existing monitoring and experience sharing tools. We finally propose a result in the form of the tool MEShaT (Monitoring and Experience Sharing Tool) and end on the perspectives offered by these researches.

## 1 INTRODUCTION

Nos recherches ont pour objectif d'offrir des outils d'aide aux apprenants et tuteurs impliqués dans des activités d'apprentissage en projet en présence ou à distance. L'apprentissage par projet est souvent appliqué pour un apprentissage complexe (i.e. qui a pour but d'amener les étudiants à acquérir et développer diverses compétences et comportements). Ce type d'apprentissage repose sur le co-développement, la responsabilité collective et la coopération [1], les apprenants étant les principaux acteurs de leur apprentissage. Ce type d'apprentissage se fait souvent en groupes encadrés chacun par un tuteur. Ces formations reposent sur des organisations riches et complexes, particulièrement pour les tuteurs et il y est difficile de mettre en œuvre des stratégies pédagogiques innovantes. La perception de l'activité de l'individu et du groupe est très difficile, surtout sans usage de Technologie de l'Information et de la Communication (TIC).

Ainsi, l'implémentation de l'apprentissage par projet dans l'enseignement supérieur ou dans un contexte professionnel reste relativement peu ambitieuse par rapport aux recommandations qui sont faites en termes d'objectifs pédagogiques [2]. De plus, les apports des théories de la cognition, comme les modèles d'apprentissage expérientiel, sont mal utilisés. Lorsque des modèles de ce type sont mis en œuvre, par exemple celui de Kolb [3], c'est souvent l'action (via l'articulation conceptualisation-expérience) qui est favorisée au détriment de la réflexion et de l'expérience personnelle [4].

Notre propos, dans une démarche bottom-up, est (1) d'observer, selon des méthodes de Knowledge Management, une formation d'apprentissage à la gestion de projet [5]. Cette formation est supportée par une organisation riche et complexe, particulièrement pour les tuteurs (cf. §2). Son observation nous permet de (2) mettre en évidence et analyser les problèmes vécus par les acteurs (étudiants, tuteurs, concepteurs) et de (3) proposer des outils capables de les résoudre ou les améliorer. Nous étudions en particulier la pertinence et les limites des outils existants concernant le suivi et le partage d'expérience. Nous proposons un résultat sous la forme de l'outil MEShaT (« *Monitoring and Experience Sharing Tool* ») et concluons sur les perspectives offertes par ces recherches.

## 2 ETUDE DE CAS : LA FORMATION A LA GESTION DE PROJET

### 2.1 Descriptif de la formation

La formation est composée (1) d'un cours qui présente théoriquement les principes et méthodes de la gestion de projet et (2) d'une application des principes théoriques dans le cadre d'un projet. Appelé Projet Collectif (PCo), il est mené par des groupes (12 groupes de 8 étudiants) qui répondent à différents besoins industriels [5]. La formation dure 6 mois et représente un investissement d'à peu près 3000 heures de travail par projet. Les objectifs pédagogiques sont l'acquisition de compétences « dures » (e.g. planifier un projet, gérer des ressources, contrôler la qualité) et « soft » (e.g. compétences sociales de collaboration et communication, considération des autres, leadership). L'équipe pédagogique (cf. figure 1) est composée d'un enseignant, de 24 tuteurs (un tuteur technique et un tuteur management par groupe), de 2 responsables (technique et management) en charge de la coordination de tous les tuteurs et d'un directeur responsable de l'organisation de la formation.

Le projet se déroule en quatre phases :
- Novembre : réponse à l'appel d'offre (formalisation des spécifications du client).
- Décembre : élaboration du plan directeur, définition des outils de conduite du projet (tableaux de bord) et des règles pour tester la qualité des livrables.
- Janvier à Mars : développement du produit ou de l'étude.
- Avant mi-Avril : rendu d'un rapport technique et d'un rapport management (analyse du déroulement et des problèmes rencontrés) et présentation théâtralisée des résultats produits et des analyses réflexives faites.

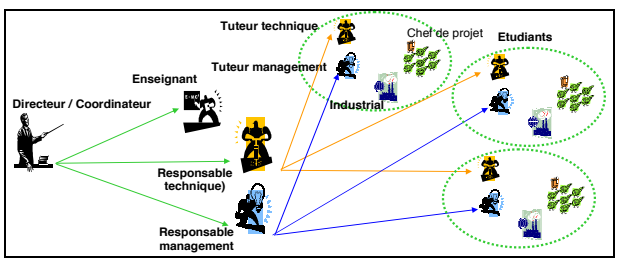

*Fig 1 : Organisation pédagogique de la formation*

L'organisation pédagogique est innovante dans la mesure où elle s'appuie sur des modèles d'apprentissage expérientiels comme celui de Berggren and Söderlund [4]. Ces derniers ont étendu le modèle de Kolb [3] basé sur quatre phases : expérience concrète, observation réflexive, conceptualisation, et expérimentation active. Ils y ont inclus des actions sociales (cf. figure 2). Les processus d'*articulation* et *réflexion* permettent l'abstraction du savoir, les processus d'*investigation*, de *mise en œuvre* et de *diffusion* peuvent être réalisés individuellement ou collectivement et contribuent à la compréhension effective des concepts.

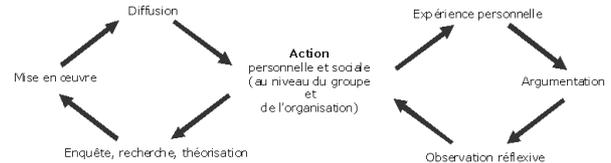

*Fig 2 : Cycle d'apprentissage étendu [4]*

Dans le cadre de la formation, l'expérience personnelle de l'étudiant évolue en suivant la double boucle. L'enseignant présente les concepts liés aux compétences « soft » et « dures ». La phase d'*articulation/réflexion* se fait lors des réunions d'équipe ou lors de débriefings hebdomadaires avec les tuteurs pour comprendre les concepts les plus simples. Les concepts complexes s'acquièrent tout au long de la formation et leur compréhension (ou la réalisation par l'élève qu'il les a acquis) se fait dans les dernières étapes du projet par la rédaction des livrables, rapports ou la présentation théâtralisée. En effet, les étudiants y présentent leurs bonnes/mauvaises pratiques et sont ainsi forcés de comprendre l'expérience qu'ils ont eux-mêmes construite dans le cadre du projet. L'expérimentation effective et le test de la bonne compréhension théorique (représentés par la boucle de gauche) sont faits dans les actions de travail en groupe projet.

Une autre caractéristique de la formation est de bien prendre en compte les bénéfices possibles des phases d'*investigation* et de *mise en oeuvre* d'une part et de *diffusion* d'autre part. En effet, les formations en management de projet s'appuient souvent sur un cas d'étude. Dans notre cas, la phase d'*investigation* est amplifiée par le fait que les projets correspondent à des cas concrets, réels et sans solution prédéfinie. Ceci implique un challenge pour l'étudiant, complexifie la phase de mise en œuvre mais améliore le processus expérientiel. Le processus de *diffusion* est réalisé soit dans le cadre du groupe via les réunions, soit à la fin du projet dans les présentations théâtralisées. Les élèves présentent leur expérience et leurs avis sur la pédagogie, les tuteurs ou leur propre formation. Ainsi, cette étape permet un partage d'expérience entre les acteurs et donne les moyens d'améliorer la formation.

Pour soutenir ce processus, les tuteurs jouent des rôles variés dépendant des compétences que les étudiants ont à acquérir [6]. Selon la taxonomie proposée par Garrot [7], pour l'acquisition de compétences « soft » les tuteurs sont des catalyseurs sociaux (créent un environ-nement convivial et incitent les apprenants à participer), des catalyseurs intellectuels (posent des questions et incitent les étudiants à discuter et critiquer), des « individualisateurs » (aident chaque étudiant à surmonter ses difficultés, à estimer ses besoins, ses difficultés, ses préférences) et des « autonomiseurs » (aident les étudiants à réguler leur apprentissage et à acquérir de l'autonomie). Pour l'acquisition de compétences « dures », les tuteurs sont des coachs relationnels (aident les étudiants à apprendre à travailler en groupe), des pédagogues (redirigent les activités du groupe dans une voie productive, clarifient des points de méthodologie, apportent des ressources), des experts du contenu et des évaluateurs. Les tuteurs encadrent un projet unique et non reproductible, la plupart du temps en face-à-face. Aucun outil de support au suivi des étudiants, à la communication, à l'organisation ou la capitalisation n'est proposé. L'appréciation des étudiants se fait de façon implicite à partir du nombre et de la qualité des interactions. En terme de communication et coordination, chaque tuteur travaille individuellement avec son groupe et ne communique pas avec l'autre tuteur du même groupe afin d'avoir une vision complète de l'activité du groupe.

Pour mieux comprendre les problèmes de cette formation, nous avons procédé à une observation.

### 2.2 Méthodologie d'observation

La méthodologie d'observation utilisée est généralement employée dans des projets de KM (*Knowledge Management*) et est une adaptation de la méthode MASK (*Method for Knowledge System Management*) [8]. Cette méthode est assimilable à un retour d'expérience et modélise les systèmes industriels complexes en s'appuyant sur les documents produits par l'organisation et des entretiens d'acteurs du terrain. Ces sources d'informations sont analysées pour identifier les concepts « produits, acteurs, activités, règles et contraintes » et les décrire précisément dans des fiches ICARE (*Information, Constraint, Activity, Rule, Entity*). L'analyse identifie aussi tous les problèmes exprimés. Ils sont, eux aussi, décrits précisément dans des fiches : les fiches RISE (*Reuse, Improve and Share Experiment*). Les fiches ICARE et RISE sont organisées dans un diagramme global qui montre toutes les interrelations [5].

Dans le cadre de cette recherche, nous avons analysé les résultats des fiches RISE. Pour les construire, nous nous somme appuyés sur les retours d'expérience de 41 élèves et 24 groupes d'élèves. Le groupe de 41 étudiants était composé de 23 hommes et 18 femmes entre 22 et 25 ans. 38 d'entre eux avaient suivi le projet l'année précédente alors que 3 suivaient la formation au moment de l'observation et étaient chef de projet. Les observations ont consisté en des REX directs sur ces 41 élèves (sous la forme d'entretiens pour les 3 élèves en formation et de séances de formalisation directes de leurs souvenirs pour les autres). Elles ont été complétées par des REX indirects réalisés par analyse des problèmes exprimés dans 24 « rapports management » (qui sont un des livrables du projet) et qui représentent l'expérience des différentes équipes. Ces REX ont mis en évidence 36 catégories de problèmes que nous présentons par fréquence d'expression.

### 2.3 Problèmes observés

Comme nous pouvons le voir sur la figure 3, les principaux problèmes identifiés concernent :

- La gestion du travail d'équipe par le groupe lui-même (57%) : manque de compétences en gestion de projet, difficulté de travailler en groupe, manque de responsabilité de la part des étudiants.
- L'activité du tuteur et son impact sur l'organisation du projet (31%) : manque de cohérence, de coordination et de communication entre tuteurs, manque de communication entre tuteurs et étudiants, manque d'information pour les tuteurs sur les objectifs pédagogiques et sur les connaissances et compétences que les étudiants ont à acquérir.
- La conception de la formation (8%) : manque de temps pour travailler, calendrier non adapté.
- Le suivi des activités individuelles et collectives et l'évaluation (4%) : difficulté pour les tuteurs d'évaluer les étudiants individuellement, manque de traçabilité des actions des étudiants, manque de discussions entre tuteurs.

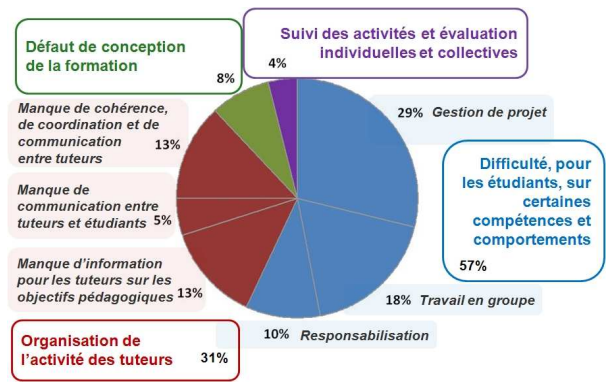

*Fig 3 : Problèmes observés dans le cadre de la formation*

Les problèmes concernant la conception et l'organisation de la formation ont été partiellement résolus mais de nombreux problèmes, principalement liés aux tuteurs, persistent. Le manque d'outils de suivi et de transfert d'expertise entraîne d'importants dysfonctionnements dans l'organisation de la formation et de ce fait une insatisfaction de la part des tuteurs et des étudiants (en particulier à propos de l'acquisition de connaissances et expertise). C'est pourquoi nous avons pour objectifs d'aider les tuteurs, d'une part, à suivre et évaluer les étudiants et le groupe et, d'autre part, à échanger des informations, à se coordonner et à développer des compétences et expertise. Bien que le contexte pédagogique soit en présentiel, nous espérons bénéficier d'outils pour supporter cette activité. Dans la section suivante, nous étudions les outils existants de suivi et de partage d'expérience.

# 3 OUTILS DE SUPPORT AUX ACTIVITES D'APPRENTISSAGE

Dans cette partie, nous présentons des outils qui aident les tuteurs à suivre les activités des étudiants et à communiquer entre eux. Nous étudions comment ces outils peuvent aider les tuteurs et régler les problèmes identifiés dans la partie précédente. Finalement, nous montrons qu'aucun de ces outils ne répond à tous les besoins et c'est la raison pour laquelle nous développons notre propre outil.

## 3.1 Outils de suivi

De nombreux outils ont été développés pour aider les tuteurs à suivre les activités *individuelles synchrones* des étudiants à distance. ESSAIM [9] donne une vue globale du parcours de l'étudiant dans le cours. Les tuteurs ont une perception de l'activité en regardant le parcours, les actions et les productions de chaque étudiant. FORMID [10] offre une interface au tuteur avec une vue globale de la classe pendant une session (e.g. les étudiants connectés, la progression dans le cours) ou un zoom sur une étape précise du cours (validée ou non par la classe, par un étudiant ou par un groupe d'étudiants afin d'identifier leurs difficultés). Ces outils fonctionnent sur un environnement synchrone avec des traces générées automatiquement. Ils ont donc été conçus uniquement pour les tuteurs et n'offrent pas aux étudiants la possibilité de réguler leurs apprentissages sur une longue période. De plus, ils ne sont pas conçus pour des situations d'apprentissage asynchrones pour lesquelles le tuteur a besoin d'informations sur les activités des étudiants sur une longue période.

D'autres outils ont été développés pour suivre les activités *asynchrones* et permettent d'accompagner les étudiants vers l'autonomie en les aidant à réguler leur apprentissage en déterminant eux-mêmes leur état de progression dans le cours. Croisières [11] offre des services qui supportent les apprenants de manière individuelle dans leurs apprentissages et les aident en situation d'autonomie. Les apprenants choisissent leurs activités d'apprentissage en fonction de leur propre stratégie d'apprentissage. Reflet [12] est un outil conçu pour montrer l'état de progression d'un étudiant ou une classe. Il fournit des informations aux tuteurs qui suivent les étudiants en apprentissage à distance et aux étudiants qui ont un retour sur leur progression en comparaison des objectifs d'apprentissage et des autres étudiants. Les apprenants déterminent eux-mêmes leur état d'avancement dans le cours en fonction des tâches qu'ils ont à réaliser et les tuteurs peuvent refuser la validation pour certaines tâches des apprenants.

D'autres outils ont été développés pour aider le tuteur à suivre les activités *collectives*, pas seulement individuelles. Par exemple, SIGFAD [13] fournit des statistiques sur les interactions pour modéliser la collaboration au sein des groupes à trois niveaux : les individus (leur productivité en terme de réalisation des activités et leur sociabilité qui indique leur niveau de communication avec les autres membres du groupe), le groupe (personnes présentes, absentes, ou endormies, l'état de réalisation des activités du groupe) et l'activité (le niveau de réalisation d'une activité par tous les participants). TACSI [14] et LCC [15] proposent plus spécifiquement une perception de l'activité individuelle au sein du groupe. TACSI [14] permet de percevoir l'activité des apprenants dans une tâche collective (leurs contributions dans des activités collectives et dans les discussions) et la perception de la situation de l'apprenant dans la dynamique du groupe (le comportement social et le statut sociométrique). Le dispositif LCC [15] permet de mesurer l'efficacité du groupe à réaliser une tâche et la qualité du travail de groupe (correspondant aux compétences clés telles que l'orientation, le leadership, le suivi, le feedback, le back-up et la coordination). Ces indicateurs sont relativement bien adaptés à notre contexte et nous les considérons pour le développement d'un outil de suivi.

Cependant, la formation qui nous intéresse n'utilise pas d'activité instrumentée et ne permet donc pas d'avoir des traces générées automatiquement, comme c'est le cas dans les outils étudiés. C'est pourquoi nous devons penser à d'autres moyens de collecter des informations sur les activités des apprenants. Nous recommandons de s'inspirer d'outils de suivi conçus pour permettre aux apprenants d'aller vers l'autonomie et de réguler leur apprentissage sur une longue période, comme Croisières [11], en les combinant avec des outils qui aident à construire une réflexion personnelle sur la pertinence des connaissances acquises et les modalités de ces acquisitions, comme le font les outils métacognitifs. La proposition d'Azevedo [16] nous semble particulièrement pertinente puisqu'elle prend en compte le point de vue des apprenants concernant leur cognition (e.g. l'activation de connaissances antérieures, la planification des apprentissage, les stratégies proposées), leur métacognition (e.g. le sentiment de savoir, le jugement sur l'apprentissage, l'évaluation du contenu), leur motivation (e.g. le jugement sur leur auto-efficacité, la valeur de leurs tâches, leur intérêt, l'effort produit) et leur comportement (e.g. démarches constructives et volontaires comme la recherche d'aide ou le traitement des taches affectées).

Tous les outils proposés dans cette partie sont exclusivement centrés sur les activités des apprenants et n'aident ni les apprenants ni les tuteurs à avoir une réflexion sur leur activité. De manière à soutenir l'activité et les réflexions non seulement des apprenants mais aussi des tuteurs, nous pensons essentiel d'avoir une base commune de structuration. L'utilisation des contrats d'apprentissages [4] qui définissent par exemple ce que les étudiants et les tuteurs ont à apprendre, comment ils sont sensés le faire, comment ils seront accompagnés et comment ils sauront qu'ils ont appris, est un bon moyen de le faire.

Enfin, tous ces outils n'aident pas les tuteurs à comprendre ou interpréter ce qu'ils observent. Ils fournissent des informations utiles aux tuteurs mais ces informations sont plus quantitatives que qualitatives et ne permettent pas d'évaluer la qualité des contributions ou productions, ou d'expliquer le comportement des apprenants. Ces outils ne peuvent être utiles aux tuteurs que s'ils savent comment les utiliser, comment interpréter les informations fournies et comment réagir efficacement au moment adéquat. Finalement, ces outils s'adressent à un tuteur individuellement et ne permettent ni de se coordonner au niveau du suivi d'un même groupe de projet ni d'échanger sur leur activité afin d'acquérir de l'expertise. C'est pourquoi nous étudions dans la partie suivante les outils permettant aux tuteurs d'échanger entre eux afin de s'entraider et ainsi développer leurs compétences.

### 3.2 Outils de partage d'expériences

Les résultats de précédentes études ont montré que les tuteurs n'ont pas d'outils adaptés pour avoir des échanges et/ou les structurer [17], comme le permettent par exemple les systèmes basés sur l'expérience ou les livres de connaissances [18]. De plus, ils ont rarement des outils pour se coordonner ou des espaces dédiés pour échanger entre pairs.

Pour compenser le manque de formation et d'aide formelle, des Communautés de Pratique (CoPs) de tuteurs émergent grâce aux technologies Web [19 ; 20]. Les tuteurs se rassemblent de manière informelle au sein de CoPs du fait qu'ils ont des pratiques, des intérêts et des buts communs (e.g. pour partager des idées et des expériences, pour construire des outils communs, pour développer des relations entre pairs) [21]. Les membres des CoPs échangent des informations, s'entraident afin de développer leurs compétences et expertise et résoudre des problèmes de façon innovante. Ils développent une identité communautaire autour de connaissances partagées, d'approches communes et de pratiques établies et créent un répertoire partagé de ressources communes [22].

En terme d'outils, les technologies Web tels que blogs, listes de diffusion, chat et mail permettent de communiquer et de capitaliser des échanges. Cependant ils sont relativement peu structurés et non contextualisés. Ils ne permettent donc pas la construction d'une connaissance réutilisable (seuls les forums apportent un niveau supérieur de structuration grâce à la représentation spatiale sous forme de flux de discussion qui montrent des relations entre les messages).

Quelques travaux tendent à résoudre ce problème en offrant aux tuteurs des outils pour des activités spécifiques. Les environnements Tapped In [22] et CoPe_it! [23] renforcent la participation et la sociabilité des membres. Tapped In offre une « maison » virtuelle alors que CoPe_it! soutient la collaboration entre pairs. D'autres outils comme doceNet [24] favorisent la création de ressources contextuelles et la recherche contextuelle. Cependant, tous ces environnements favorisent soit la sociabilité au détriment de la réification des ressources produites, soit l'accumulation et l'indexation de ressources contextualisées au détriment de la sociabilité et de la participation des membres.

La plate-forme communautaire TE-Cap [22] supporte une bonne structuration de l'information sans contraindre la participation des membres (par exemple concernant la forme de la communication). En effet, les tuteurs peuvent échanger par le biais de forums contextualisés : ils associent des tags aux discussions pour en décrire le contexte. Ces tags sont des rubriques d'une taxonomie du tutorat, présentée de façon interactive et évolutive (les tuteurs peuvent proposer de nouvelles rubriques à la taxonomie). Cette plate-forme, associée à un outil de suivi, peut répondre aux besoins d'acquisition et capitalisation des connaissances et compétences sur la réalisation de l'activité des tuteurs et sur l'utilisation des outils de suivi.

## 4 MEShaT : UNE PLATE-FORME POUR TUTEURS ET ETUDIANTS

Nous avons pu voir qu'il n'existe pas d'outil qui réponde à l'ensemble de nos besoins. Nous proposons donc une nouvelle plate-forme nommée MEShaT (cf. figure 4), qui se fonde sur les outils retenus dans la partie précédente pour proposer une solution globale. Elle fournit des informations pour suivre les activités et des outils pour supporter l'acquisition et le transfert d'expertise. MEShaT propose trois interfaces personnalisées selon que l'utilisateur est un étudiant, un groupe de projet ou un tuteur. Chaque interface consiste en (1) un outil de suivi (sous la forme d'un tableau de bord) qui aide l'acteur concerné à avoir une vue globale de son activité et (2) un outil de publication qui lui permet de partager son expérience. Trois tableaux de bord (TDB) sont proposés : deux sont dédiés aux étudiants pour les aider à gérer d'une part l'avancement de leur projet et d'autre part l'évolution de leurs propres apprentissages, l'autre est dédié aux tuteurs pour les aider à suivre l'activité des élèves et leurs processus d'apprentissage.

Le TDB de suivi du projet est assimilable à un outil de gestion de projet dédié au groupe et montre différents indicateurs représentatifs de l'évolution de l'état d'esprit du groupe (motivation, satisfaction, relation avec le client), de la planification du travail (diagramme de Gantt, activités à réaliser, retards, délais) et de l'activité effective des membres (temps de travail, livrables finalisés). Ce TDB est dédié au chef de projet pour qu'il organise son management mais aussi aux membres de l'équipe pour qu'ils expriment leurs points de vue personnels sur le travail de groupe et qu'ils situent leur activité par rapport à celle des autres membres. Ainsi, il fournit une partie des informations nécessaires à la réalisation du processus métacognitif décrit ci-dessous et à la gestion de l'activité des élèves et des groupes par le tuteur.

Le TDB métacognitif prend en compte le point de vue individuel des élèves sur leur cognition, leur métacognition, leur motivation et leurs comportements afin de construire des indicateurs réflexifs qui les aident lors du processus d'apprentissage, en particulier pour une évolution de comportement [16]. La cognition est prise en compte sous la forme de niveaux de compétences que chaque étudiant auto-évalue toutes les semaines par rapport aux objectifs pédagogiques (« soft » et « hard ») ou aux objectifs de projet fixés. La métacognition représente les impressions des élèves à propos de leurs compétences et de leurs processus d'apprentissage (sur le plan du niveau, de la forme, du contexte, de l'appréciation). Les élèves décrivent en particulier leur motivation, jugent leur efficacité, l'intérêt des taches qu'ils ont à faire et l'effort qui est requis. Ils formalisent leur comportement en expliquant les actions qu'ils ont mis en œuvre pour atteindre les objectifs ou résoudre les difficultés. Ces informations, présentées de manière synthétique et visuelle dans un TDB, les aident à construire leur propre point de vue sur la connaissance construite ou sur le nouveau comportement acquis ainsi que sur la pertinence des stratégies qu'ils ont mises en œuvre. Ils apprennent ainsi à apprendre et à évoluer dans le cadre de leurs futures activités.

Le TDB des tuteurs leur donne des informations sur les activités individuelles et collectives telles que l'orientation du groupe, le leadership, le suivi, le feed-back et la coordination [15]. Ces indicateurs sont construits sur la base des informations données par les élèves dans les TDB de suivi de projet et TDB de suivi individuel décrits précédemment. Ainsi les tuteurs ont plus de facilité à intervenir durant le projet ou à comprendre *a posteriori* les processus individuels et collectifs de travail.

Les outils de publications sont utilisés pour expliciter les contextes de réalisation des actions et transférer l'expérience. Ils sont constitués de :

- Blogs (un par étudiant et par groupe) : espaces où les étudiants peuvent décrire librement par exemple le contexte de réalisation de leurs actions et leur état d'esprit. Ces blogs aident les membres du groupe et les tuteurs à comprendre le contexte du projet, la valeur de certains indicateurs (tels que les délais où l'état d'esprit du groupe) et à anticiper ou résoudre plus rapidement des problèmes.
- TE-Cap [22] : fourni aux tuteurs pour permettre l'émergence d'une Communauté de Pratique (CoP) constituée de tous les tuteurs de la formation. Le modèle d'indexation est construit sur trois sujets principaux : (1) leurs rôles et tâches, (2) le calendrier du projet (pour se coordonner) et (3) la progression spécifique de chaque groupe. En échangeant, les tuteurs peuvent acquérir de l'expertise sur leurs rôles et des connaissances sur leur application sur le terrain. TE-Cap peut être considéré comme un outil de transfert d'expertise.

Une partie fixe et transversale aux trois TDB montre des informations accessibles par tous les acteurs : le contrat d'apprentissage [4] et l'agenda. Le contrat d'apprentissage aide chaque acteur à s'évaluer par rapport au modèle pédagogique et à mieux atteindre ses objectifs.

L'originalité de Meshat est de pouvoir combiner dans une plate-forme unifiée personnalisable trois types de supports de formation proposés, comme nous l'avons vu dans l'état de l'art, la plupart du temps de manière unitaire : la *supervision* (via les TDB), *l'information et la communication* (via des actions de publication dans les blogs et TE-Cap) et la *coordination* (via les agendas et le contrat d'apprentissage). Cette combinaison est nécessaire pour répondre aux problèmes identifiés dans la formation et rend les effets attendus de chaque support plus efficaces.

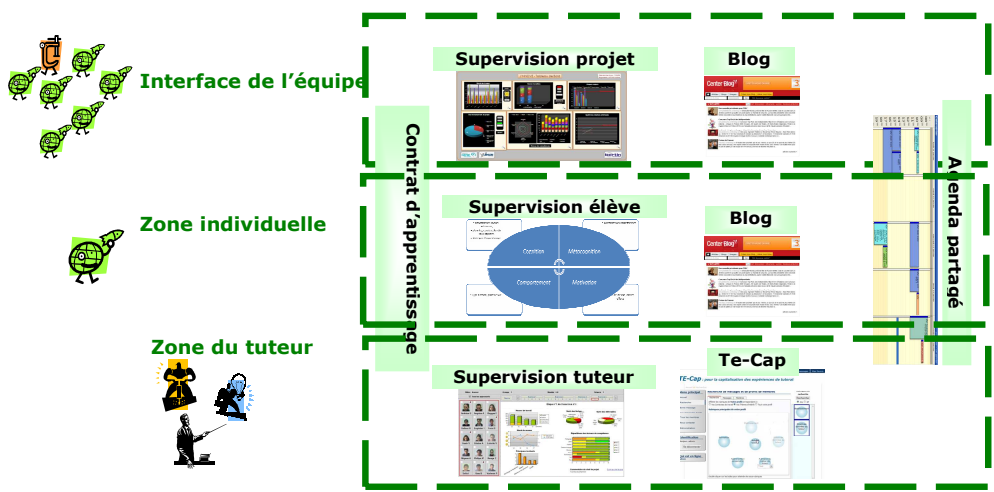

*Fig. 4 : MEShaT : Outil de suivi et de partage d'expérience pour l'apprentissage par projet*

En effet, du point de vue des apprenants, la possibilité d'avoir une supervision individuelle comme dans [11] combinée à une supervision du groupe comme dans [14] et [13], facilite la prise en compte de l'apprenant au sein du groupe, le travail de groupe, sa cohésion et le professionnalisme des étudiants (en rendant plus tangibles les conséquences de leurs actes et en les informant). La combinaison de la supervision individuelle (outil métacognitif), du contrat d'apprentissage et des blogs aide les étudiants de manière plus complète que dans [11] par exemple à construire leurs connaissances (par la prise d'information et l'échange entre pairs) et à renforcer leur motivation (par une meilleure compréhension de ce qu'ils ont à faire et pourquoi ils le font).

Si ces phénomènes n'apparaissent pas naturellement, MEShaT va permettre aux tuteurs d'avoir une aide pour les faire émerger. En effet, MEShaT renforce le lien tuteur-étudiant en assurant un suivi continu du processus d'acquisition de connaissances alors que les applications présentées précédemment se concentrent plutôt sur le suivi d'activité [9][10][12]. L'interface tuteur les aide aussi à assumer certains de leurs rôles, tels que celui de coach relationnel, de catalyseur social (concernant le travail de groupe et le leadership), de catalyseur intellectuel (en posant des questions précises et conceptualisées pour inciter les étudiants à discuter et poser des questions critiques), d'expert et de pédagogue alors que les applications de supervision que nous avons pu voir précédemment [10][12] se concentrent sur leur rôle d'évaluateur. De plus, l'association de TE-Cap [22] avec le contrat d'apprentissage offre aux tuteurs un espace pour affiner ou développer leur expertise de manière plus globale que l'outil utilisé individuellement.

## 5 CONCLUSION

Dans le cadre de cet article nous nous sommes attachés à étudier comment les méthodes de KM et du Web 2.0 peuvent être utilisées pour améliorer l'apprentissage en face-à-face ou à distance, dans le cadre d'activités en mode projet. Nous avons proposé une solution qui intègre des outils de suivi de l'activité et de partage d'expériences pour les tuteurs et apprenants. Afin de mieux comprendre les besoins de ces acteurs dans ce contexte, nous avons étudié une formation spécifique à la gestion de projet. La spécificité de cette formation réside sur la mise en œuvre d'un modèle d'apprentissage expérientiel sous la forme de pédagogie par projet. De plus, les objectifs pédagogiques sont complexes (formalisés sous la forme de compétences « dures » et « soft ») et le modèle s'appuie sur une organisation sociale riche et variée qui considère plusieurs types de tuteurs, des clients industriels avec des projets industriels concrets.

Dans la première partie de cet article, nous avons décrit précisément cette formation et expliqué comment les modèles de l'apprentissage expérientiel sont concrètement mis en œuvre. Ensuite, nous avons exposé les résultats d'une étude relative aux problèmes observés dans ce cadre. Elle met en évidence des défauts de coordination, de supervision et de transfert d'expériences entre tuteurs, apprenants ou tuteurs/apprenants, ce qui engendre de multiples dysfonctionnements. Dans la troisième partie, nous avons présenté les outils actuellement réalisés pour supporter ce type de processus. L'analyse qui en découle montre qu'il n'existe pas d'outil qui aide à la fois les tuteurs et les apprenants et qu'il n'y a pas de stratégie clairement définie pour acquérir, transférer ou capitaliser l'expérience des acteurs. L'outil que nous proposons dans la quatrième partie, MEShat, vise à pallier ces manques. Il se présente sous la forme d'un tableau de bord personnalisé selon que l'utilisateur est un apprenant, un groupe projet ou un tuteur. Les différentes interfaces présentent des indicateurs, des outils de publication (sous la forme de blogs et CoP) et des outils de coordination.

L'originalité de cette solution repose sur le fait qu'elle favorise, en terme de modèle d'apprentissage, l'articulation entre action (expérience ou conceptualisation) et réflexion. Cette approche améliore donc l'acquisition de compétences complexes (e.g. de gestion, communication et collaboration) qui requière une évolution de comportement. Nous avons pour but de rendre les étudiants capables « d'apprendre à apprendre » et d'évoluer selon les contextes. Nous facilitons leur capacité à avoir une analyse critique de leurs actions selon les situations qu'ils rencontrent. Nous sommes en cours de réalisation de la plate-forme et nos recherches futures vont consister à tester MESHaT sur une longue période afin de valider expérimentalement nos hypothèses. Nous allons également observer comment les acteurs s'approprient les technologies et comment celles-ci participent à la redéfinition de leurs rôles.